\documentclass[aps,pre,twocolumn,showpacs,floatfix,superscriptaddress]{revtex4-2}
\usepackage{subfigure,amsmath, amsthm, amssymb}
\usepackage{graphicx}
\usepackage{graphics}
\usepackage[export]{adjustbox}
\usepackage[usenames]{color}
\usepackage{float}
\usepackage{physics}
\usepackage{tabularx,ragged2e}
\usepackage[geometry]{ifsym}
\usepackage{wrapfig,lipsum,booktabs}
\newcommand{\be}{\begin{equation}}
\newcommand{\ee}{\end{equation}}
\newcommand{\bea}{\begin{eqnarray}}
\newcommand{\eea}{\end{eqnarray}}

\begin{document}
\title{Dissipation and recovery in collagen fibrils under cyclic loading: a molecular dynamics study}
\author{Amir Suhail}
\email{amirs@imsc.res.in}
\affiliation{The Institute of Mathematical Sciences, CIT Campus, Taramani, 
Chennai 600113, India}
\affiliation{Homi Bhabha National Institute, Training School Complex, Anushakti Nagar, Mumbai 400094, India}
\author{Anuradha Banerjee}
\email{anuban@iitm.ac.in}
\affiliation{Department of Applied Mechanics, IIT-Madras, Chennai-600036, India}
\author{R. Rajesh}
\email{rrajesh@imsc.res.in}
\affiliation{The Institute of Mathematical Sciences, CIT Campus, Taramani, 
Chennai 600113, India}
\affiliation{Homi Bhabha National Institute, Training School Complex, Anushakti Nagar, Mumbai 400094, India}
  
\date{\today}

\begin{abstract}
The hysteretic behavior exhibited by collagen fibrils, when subjected to cyclic loading, is known to result in both dissipation as well as accumulation of residual strain. On subsequent relaxation, partial recovery has also been reported. Cross-links have been considered to play a key role in overall mechanical properties. Here, we modify an existing  coarse grained molecular dynamics model for collagen fibril with initially cross-linked collagen molecules, which is known to reproduce the response to uniaxial strain, by incorporating reformation of cross-links to allow for possible recovery of the fibril. Using molecular dynamics simulations, we show that our model successfully replicates the key features observed in experimental data, including the movement of hysteresis loops, the time evolution of residual strains and energy dissipation, as well as the recovery observed during relaxation. We also show that the characteristic cycle number, describing the approach towards steady state, has a value similar to that in experiments. We also emphasize the vital role of the degree of cross-linking on the key features of the macroscopic response to cyclic loading. 
\end{abstract}

\maketitle

\section{Introduction}

Collagen protein, the organic matrix of many biological tissues such as bones, tendons, ligaments, cartilage, skin, and cornea etc., depending on the function of the tissue, provides desirable structural stability, elastic resilience, strength and toughness~\cite{Kalder2007, Fratzl2008,Shoulders2009,hulmes2008collagen}. This versatility as a building material is a direct consequence of mainly the variations in the hierarchical structure of the protein and also the tuneable degrees of mineralization. The hierarchical structure of collagenous tissues has the basic building block of collagen fibrils which self-assemble into different tissues with wide ranging properties~\cite{bose2022_review,yang2019_review,OTTANI2002,FRATZL20071263}. Understanding the structure-property relationship of collagen at different length scales and developing predictive models is important not only for identifying the origins of the diversity of properties observed in different tissues but also for achieving important clinical objectives like health assessment of tissues and optimizing therapeutic procedures~\cite{bose2022_review,rezvani2021collagen,lee2001biomedical}.


Mechanical response of collagen has been experimentally probed at multiple length scales, both under monotonic tensile as well as cyclic loading.To determine the mechanical response at the smallest length scales (around $300 nm$), atomic force microscopy (AFM) and optical tweezer experiments have been conducted~\cite{Thompson2001,SUN2002,SUN2004,BOZEC2005}. AFM experiments on collagen from bovine Achilles tendon have revealed a force-extension pattern characterized by multiple sharp drops in force as extension increases, resembling a saw-tooth shape~\cite{Thompson2001}. These drops have been attributed to the breaking of sacrificial bonds, possibly cross-links, which release hidden lengths. When a time delay is introduced between cycles, partial recovery is observed, suggesting reformation of some of the sacrificial bonds~\cite{Thompson2001}. Similar force-extension curves have also been observed in the unfolding of titin~\cite{Reif1997}, a biological polymeric adhesive found in nacre~\cite{Smith1999}, etc. At the length scale of collagen molecules, X-ray diffraction has directly linked sacrificial bonds or cross-links to the folding back of the molecule at the C-terminal telopeptides~\cite{ORGEL2001}.

Collagen molecules assemble in a staggered manner to create long collagen fibrils that are micrometers in length and have diameters ranging from tens to hundreds of nanometers~\cite{Fratzl2008}. This staggered arrangement gives rise to distinct D-periodic banding patterns along the length of the fibril, with regions of gap and overlap~\cite{PETRUSKA1964,Orgel2006}. The structure of the fibril is stabilized by enzymatic covalent cross-links (ECLs) that form between molecules at the non-helical ends (telopeptides)~\cite{Light1980,KNOTT1998,Reiser1992}. Other types of cross-links also develop within the fibril, specifically Advanced Glycation Endproducts (AGEs), which occur as a result of aging and diabetes~\cite{GAUTIERI201489}.  The degree of cross-linking, including both ECLs and AGEs, between tropocollagen molecules has been shown to significantly impact the material's response~\cite{SVENSSON20132476,SHEN2010,fessel2014advanced,GAUTIERI201795,kamml2023influence}. The breaking of these cross-links and the resulting release of hidden length can be correlated with the plateaus observed in the force-extension curve. 


Collagen experiences repetitive loads during normal bodily movements. The response of collagen fibrils to cyclic loads, including dissipation and recovery, has been investigated in in-vitro experiments, ranging from individual fibrils~\cite{SHEN2008cyclicload,SVENSSON2010,LIU2018cyclicload,van2006micromechanical} to larger tissue samples~\cite{sellaro2007effects,veres2013repeated,veres2013cross,BOSE2020_cyc_coll_film,susilo2016collagen}. In a study by Shen et al.~\cite{SHEN2008cyclicload}, fatigue tests were conducted on isolated collagen fibrils, revealing four distinct stress-strain responses: linear until failure, perfectly plastic, perfectly plastic with strain hardening, and nonlinear strain softening. All fibrils exhibited significant hysteresis (energy loss) and a residual strain (strain when no force is applied). The amount of time spent at zero force influenced the recovery of the residual strain. Similar observations have also been made in experiments involving tissue samples~\cite{veres2013cross,BOSE2020_cyc_coll_film}.

The focus of this paper is to model a cyclic loading experiment conducted by Liu et al.~\cite{LIU2018cyclicload}, in which individual collagen fibrils obtained from calf skin were subjected to repeated loading cycles under displacement control. The fibrils underwent 20 loading cycles up to a predetermined stretch ratio, $\lambda_{\max}$, and were then unloaded until reaching zero force. After the first 10 cycles, the fibrils were given a 1-hour relaxation period. The stress-stretch response of the fibrils exhibited moving and diminishing hysteresis loops, which were accompanied by accumulation of residual strains. Subsequently,  the collagen fibrils also demonstrated a recovery in both residual strain and the ability to dissipate energy when allowed to relax at zero force. These observations led to the hypothesis that reformable sacrificial bonds within the fibrils may be responsible for these characteristics. Furthermore, the cyclically loaded fibrils exhibited greater strength and toughness compared to fibrils subjected to monotonic loading. This enhancement was believed to be due to permanent molecular rearrangements, although the specific mechanism behind these improvements was not fully understood.

Different approaches have been used to model the macroscopic response of fibrils under cyclic loading. Recently, we proposed a kinetic model for collagen fibrils that takes into account the presence of hidden loops, stochastic fragmentation, and the reformation of sacrificial bonds~\cite{suhail2022kinetic}. The kinetic model successfully replicated the key features observed in experimental data~\cite{LIU2018cyclicload}, including the movement of hysteresis loops, the time evolution of residual strains and energy dissipation, and the recovery observed during relaxation. We demonstrated that the approach towards reaching a steady state is influenced by a characteristic cycle number for both residual strain and energy dissipation, and our findings were consistent with the experimental data of Ref.~\cite{LIU2018cyclicload}. Within a continuum mechanics approach,  a constitutive model was proposed that accounts for both viscoelastic and plastic deformations~\cite{fontenele2023understanding}. The model parameters were fitted using experimental data, and the experimental phenomena  was well-reproduced by the model. The model predicts plastic deformation, and improved performance after relaxation.

Both modeling approaches discussed above are at the macroscopic level. In the kinetic model, there is no spatial degrees of freedom. In the viscoelastic-plastic model, constitutive behavior are assumed for the model. In this paper, we approach the problem of cyclic loading from the microscopic point of view using molecular dynamics (MD) simulations. Coarse grained models, where the parameters have been obtained from atomistic simulations, have been used earlier to study the mechanical properties of collagen fibril.  Within,  idealized two-dimensional representation of the collagen fibril, it was shown that the fibril was able to withstand  large deformations without catastrophic failure possibly due to stretching, sliding and rupture of cross-links~\cite{BUEHLER2007}. A three dimensional model of the fibril~\cite{DEPALLE20151} incorporating enzymatic cross-links in their physiological locations and other more realistic aspects of collagen, was able to reproduce the experimentally observed~\cite{SVENSSON20132476} three-phase stress-strain response. The effect of AGE cross-links on improving the mechanical properties of fibril has recently been studied using a similar three dimensional model for the fibril~\cite{kamml2023influence}. Using a slightly modified model of the fibril in Ref.~\cite{DEPALLE20151}, it was shown that degradation of the properties of the cross-links at the fibril surface changes the mechanical properties quite drastically~\cite{MALASPINA2017549}. While cyclic loading of collagen molecules have been studied using fully atomistic models~\cite{milazzo2020mechanics,milazzo2020wave,zitnay2020accumulation}, a similar study does not exist for collagen fibrils.

In this paper, we explore the ability of existing MD models~\cite{DEPALLE20151,MALASPINA2017549,kamml2023influence} to account for the experimental features observed during cyclic loading. Additionally, we incorporate cross-link reformation into our model and evaluate its potential to explain experimental findings such as recovery upon relaxation and increased strength that may result from reformation and re-organization of cross-links. We show that with these additional features, we are able to reproduce nearly all the features of the experimentally observed~\cite{LIU2018cyclicload} macroscopic response of collagen fibrils subjected to cyclic loading and relaxation.

\section{Model}
\label{model}
\subsection{Geometry of the collagen fibril model}

We first describe the geometric details of coarse-grained fibril model and how cross-links or sacrificial bonds that break and reform are incorporated into the model. The model is a modification of existing three dimensional coarse grained models for collagen fibrils in the literature~\cite{DEPALLE20151,MALASPINA2017549}.

\begin{figure}
	\subfigure[]{\includegraphics[width=1.0\linewidth]{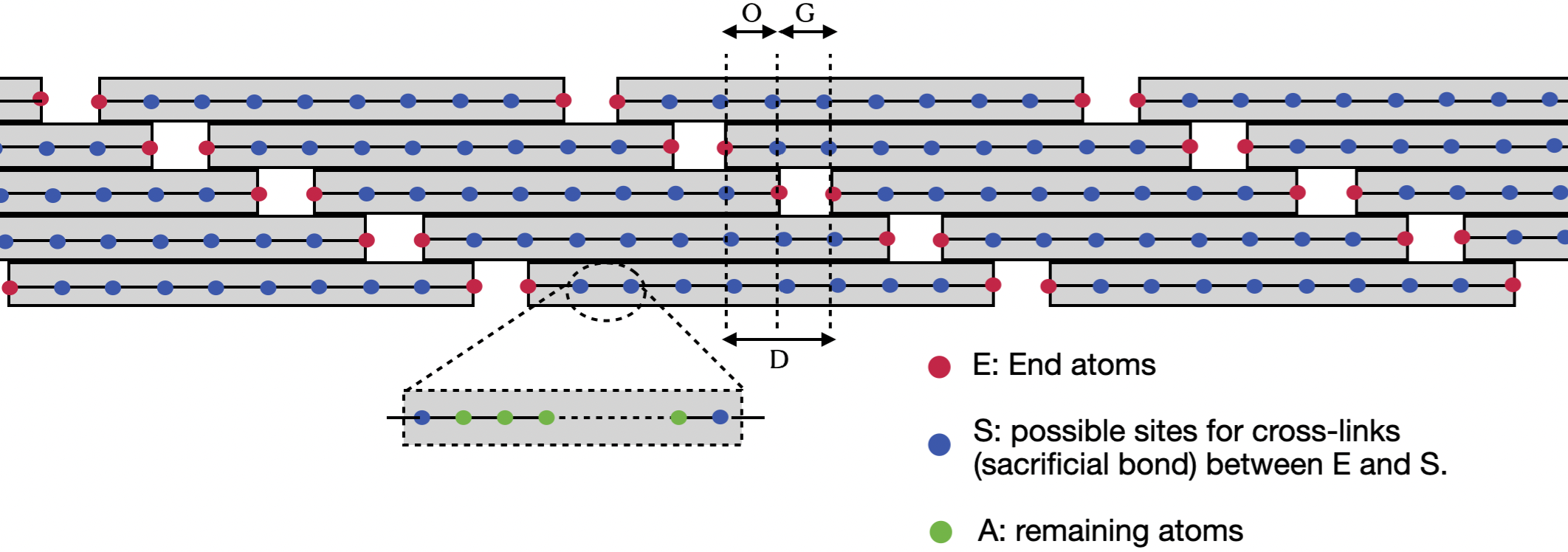}\label{fig:figure_a1}}
	\subfigure[]{\includegraphics[width=1.0\linewidth]{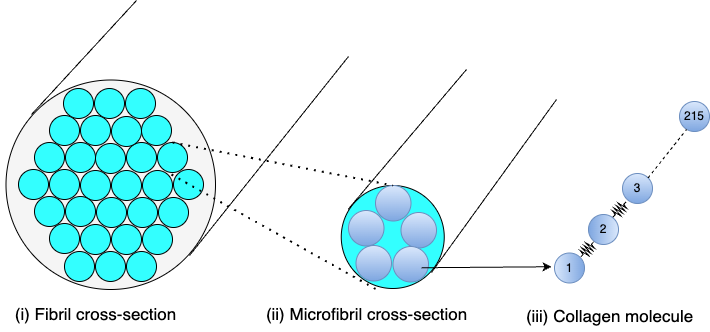}\label{fig:figure_b1}}
	\caption {(a) Schematic diagram showing the longitudinal arrangement of collagen molecules in a microfibril. Gap and overlap region represented by 'O' and 'G' respectively and D-period is shown by 'D'. (b)  Cross-sectional view of hierarchical structure of the fibril: (i) fibril, (ii) microfibril, and (iii) molecule.}
	\label{fig1: microfibril_longitudnal_schematic}
\end{figure}

A collagen molecule is represented by a linear bead-spring polymer of 217 beads, as shown in Fig.~\ref{fig1: microfibril_longitudnal_schematic}(b(iii)). The distance between two consecutive beads is $b \approx 1.4 nm$. In a microfibril, five collagen molecules  are arranged parallel to each other but in a staggered fashion longitudinally, as shown in Fig.~\ref{fig1: microfibril_longitudnal_schematic}(a),  and in a pentagonal geometry along transverse direction, as shown in Fig.~\ref{fig1: microfibril_longitudnal_schematic}(b(ii)). The diameter of a single microfibril is $\approx 3.5 \;nm$. 37 of these microfibrils, arranged in hexagonal closed packing, represents a collagen fibril, as shown in Fig.~\ref{fig1: microfibril_longitudnal_schematic}(b(i)). The staggered arrangement of collagen molecules results in a repeating gap and overlap region which give rise to the characteristic D-period ($67 \; nm$) of the collagen fibril.

It is known that collagen molecules self-assemble into fibril structure and end regions of collagen molecules (called telopeptides) forms enzymatic cross-links with their neighboring molecules to further stabilize the structure. This indicates that the end regions must have a high affinity for specific regions on the collagen molecule in order for self assembly to result in fibril formation with a precise D period. In line with this concept, we consider three types of beads ($E$, $S$ and $A$) in each collagen molecule, as shown in Fig.~\ref{fig1: microfibril_longitudnal_schematic}(a). `$E$' (shown in red) represents the end beads of each collagen molecule, `$S$' (shown in blue) represents the possible sites for cross-link formation ($E$-$S$ sacrificial bond) and `$A$'  (not shown) represents the remaining atoms in each molecule. In our model, the location of potential cross-linking sites (`$S$' beads) is chosen such that self-assembly could in principle lead to a fibril structure with a $67 \;nm$ ($D$) periodicity. We know that the diameter or bond length of the collagen polymer is $\approx1.4\; nm$. Based on this, the ratio of the $D$-period to the bond length $(D/b)$ is calculated to be approximately 48. Therefore, 48 bonds correspond to a single $D$ period. In addition, the arrangement of collagen molecules in a microfibril indicates that the second, third, fourth, and fifth molecules are staggered by $D$, $2D$, $3D$, and $4D$, respectively. As a result, beads with indices 49, 97, 145, and 193 are classified as $S$ type. We have also chosen the gap and overlap regions as $G=O=0.5D$, which is equivalent to $24$ bond lengths ($b$). Consequently, beads with indices $25, 49, 73, 97, 121, 145, 169,$ and $193$ in each collagen molecule are considered to be type $S$, as shown by the blue beads in Fig.~\ref{fig1: microfibril_longitudnal_schematic}(a). 

\subsection{Interaction potential and parameters:}

The interaction potentials and parameters closely follow that of Refs.~\cite{DEPALLE20151,MALASPINA2017549}. The interaction between all directly bonded beads including backbone of collagen molecules and the cross-links is given by a biharmonic potential as:
\begin{equation}
	F_{bond}= - \pdv{U_{bond}}{r} =  
	\begin{cases} 
		k_{T0}(r-r_0), & r  < r_1, \\
		k_{T1}(r-r_0),  &  r_1\leq r < r_{break}, \\
		0, & r> r_{break}. 
	\end{cases}
\end{equation}
where $r_0$ is the equilibrium distance between two beads, $r_1$  is the hypercritical distance, $r_{break}$ is the bond-breaking distance and $k_{T0}$ and $k_{T1}$ are spring constants. The bending interaction between triplet of consecutive beads along the backbone of each collagen molecule is given by a bending potential:
\begin{equation}
	U_{\theta} = \frac{1}{2}k_{\theta}(\theta-\theta_0)^2,
\end{equation}
where $k_{\theta}$ is bending strength and $\theta_0$ is equilibrium angle. There is no bending interaction in triplets that includes the cross-linking beads ($S$-type).
The interaction between all non-bonded beads is given by Lennard-Jones potential as:
\begin{equation}
	U_{LJ}= 4 \epsilon \left[ \left( \frac{\sigma}{r} \right)^{12} - \left( \frac{\sigma}{r} \right)^{6} \right], \quad r< r_c=2.5 \sigma,
\end{equation}
where $r$ is the distance between beads, $\epsilon$ is the strength of the potential (energy parameter), $\sigma$ is the diameter of monomer and $r_c$ is the cut-off distance. The numerical values of the different parameters are given in Table~\ref{table:t1}. 
\begin{table}
	\centering
	\caption{\label{table:t1} The parameters for the different interaction potentials of the coarse-grained MD-model of collagen fibril. }
	\begin{ruledtabular}
	\begin{tabular}{p{7.7 cm} c}
		Model parameters & Value \\
		\hline
		$\epsilon$- LJ energy parameter ($kcal \;mol^{-1}$) & 6.87 \\
		$\sigma_1$- LJ distance parameter for all non-bonded beads (except E-S pairs)(\AA) & 14.72 \\
		$\sigma_2$- LJ distance parameter for E-S pairs (\AA) & 10.0 \\
		$\theta_0$- Equilibrium bending angle ($degree$) & 180\\
		$k_{\theta}$- Bending strength constant ($kcal \;mol^{-1}\;rad^{-2} $) & 14.98 \\
		$r_0$- Equilibrium distance (tropocollagen) [\AA] & 14.00 \\
		$r_1$- critical hyperelastic distance (tropocollagen) [\AA] & 18.20  \\
		$r_{break}$- bond breaking distance (tropocollagen) & 21.00 \\
		$k_{T0}$- Stretching strength constant (tropocollagen) [$kcal \;mol^{-1}$\AA$^{-2} $] & 17.13 \\
		$k_{T1}$- Stretching strength constant (tropocollagen) [$kcal \;mol^{-1}$\AA$^{-2}$] & 97.66 \\
		$r_0$- Equilibrium distance (divalent cross-link) [\AA] & 10.00 \\
		$r_1$- critical hyperelastic distance (divalent cross-link) [\AA] & 12.00  \\
		$r_{break}$- bond breaking distance (divalent cross-link) [\AA]& 14.68 \\
		$k_{T0}$- Stretching strength constant (divalent cross-link) [$kcal \;mol^{-1}$\AA$^{-2} $] & 0.20 \\
		$k_{T1}$- Stretching strength constant (divalent cross-link) [$kcal \;mol^{-1}$\AA$^{-2}$] & 41.84 \\
		$m$- mass of tropocollagen bead [$a.m.u$] & 1358.7 \\
	\end{tabular}
	\end{ruledtabular}
\end{table}

To implement reformation, we consider `$E$' and `$S$' type atoms as special atoms. When a pair of  `$E$' and `$S$' atoms approach closer than a certain distance say $r'$, a cross-link or bond can form between them, provided neither of the participating beads is part of any existing cross-link. Once a bond is formed, the new $E$-$S$ bond is assigned the same parameters as those of a divalent cross-link. We set $r'=14$\AA, which is smaller than $r_{break}=14.68$\AA and LJ distance parameter for non-bonded $E$-$S$ pairs to be ($\sigma_2=10$\AA). In earlier models~\cite{DEPALLE20151,MALASPINA2017549} LJ parameter ($\sigma_1=14.72$\AA) was the same for all non-bonded pairs. If the $E$-$S$ bond had length $\sigma_1$, two non-bonded atoms will typically not come closer than $\sigma_1=14.72$\AA, which is larger than both the bond breaking distance ($r_{break}=14.68$\AA) for divalent cross-link and also $r'$. The equilibrium distance $r_{min}=2^{\frac{1}{6}}\sigma_1=16.52$\AA is also greater than $r_{break}$.  For bond reformation to take place, the distance between non-bonded $E$-$S$ pairs must be less than or equal to $r'$. With $\sigma_2$,  $r_{min}$ is $11.22$ \AA, allowing for the reformation of $E$-$S$ bonds.

\subsection{\label{sim_prot} Simulation protocol:}

The simulation were performed using LAMMPS \cite{LAMMPS}. Time step was set to $10\;fs$. The fibril model used in the simulations had an initial cross-link percentage of $\beta$, which indicates the fraction of  end molecules that are cross-linked. Periodic boundary conditions with a box size of length $(L+G)$ is used to mimic the fibril of infinite length with alternative gap and overlap regions, where $L$  is length of collagen molecule. The system is first equilibrated for $20\;ns$ at zero pressure in the NPT ensemble with Nose-Hoover thermostat and barostat settings of 298 K and 0 Pa, respectively. The relaxation times for the thermostat and barostat are fixed to 1 and 10 ps. A constant strain rate of $10^7\;s^{-1}$ was then applied along fibril length, and the equations of motion were integrated with a Langevin thermostat using a drag coefficient of $1\;ps$. For the cyclic loading simulation, the box was deformed up to a fixed strain ($\lambda_{\max}$) and then the direction of the applied strain rate was reversed to continue deformation until the force reached zero, which completed one loading cycle.

\section{Results}

\subsection{Uniaxial and cyclic loading of fibril model with no-reformation:}

\begin{figure}
	\centering
	\includegraphics[width = 1.0 \linewidth]{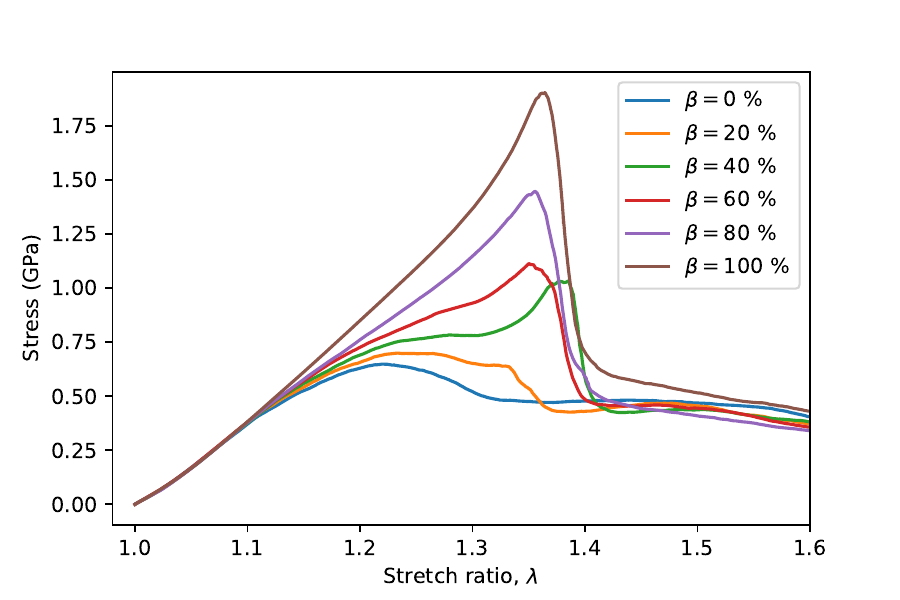}
	\caption {The stress-strain response of the collagen fibril model to applied uniaxial strain for different percentages of cross-links ($\beta$) present.}
	\label{fig2: ss_mono_no_reform}
\end{figure}

We first examine the macroscopic behavior of the coarse-grained model and its dependence on the extent of cross-linking, under monotonically increasing applied strain. The box was subjected to a constant uniaxial strain rate of $10^7\;s^{-1}$ along the longitudinal direction ($z$-axis). Initially, a fraction $\beta$,  out of the maximum possible cross-links allowed, were created. For this analysis, cross-links, once broken, were not allowed to reform for benchmarking with earlier studies. The stress-strain response observed for different $\beta$, shown in Fig.~\ref{fig2: ss_mono_no_reform}, shows an initial linear behavior, followed by a non-linear regime, and a final sharp drop after peak stress. The initial linear regime is independent of $\beta$, while the non-linear regime, depending on $\beta$, either shows shows a combination of hardening and softening regimes. When the extent of cross linking is large, the response is predominantly hardening, whereas for low extent of cross linking, it is predominantly softening. Further, we observe that for large strains, the stresses are independent of $\beta$. These observations are consistent with earlier results.

\begin{figure}
	\centering
	\includegraphics[width = 1.0 \linewidth]{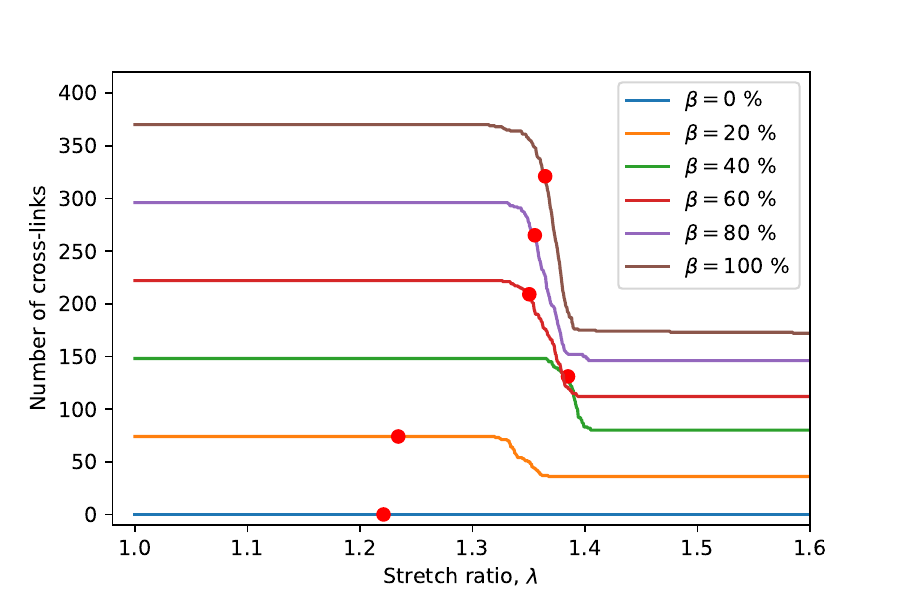}
	\caption {The number of cross-links at a given uniaxial strain  for different percentages of cross-links ($\beta$) present. The red dots represent stretch ratios corresponding to peak stress in Fig.~\ref{fig2: ss_mono_no_reform}.}
	\label{fig3: nb_mono_no_reform}
\end{figure}

The key features of the macroscopic response can be better understood in terms of the number of the intact cross-links at any given strain (see Fig.~\ref{fig3: nb_mono_no_reform}). For small strains, the number of cross-links shows no noticeable change until the strain corresponding to peak load (see Fig.~\ref{fig2: ss_mono_no_reform}) is reached. Further deformation results in a sharp decrease in the number of intact cross-links which correlates well with the sharp decrease in the stresses seen earlier. For even larger deformations, the number of cross-links again do not noticeably change with strain, though the stationary values depend on the given $\beta$, even though the stresses were seen to be independent of $\beta$ at these strains. This can be attributed to the resistance offered by the chains sliding past each other to be similar irrespective of the remaining intact cross-links.

\begin{figure}
	\centering
	\includegraphics[width = 1.0 \linewidth]{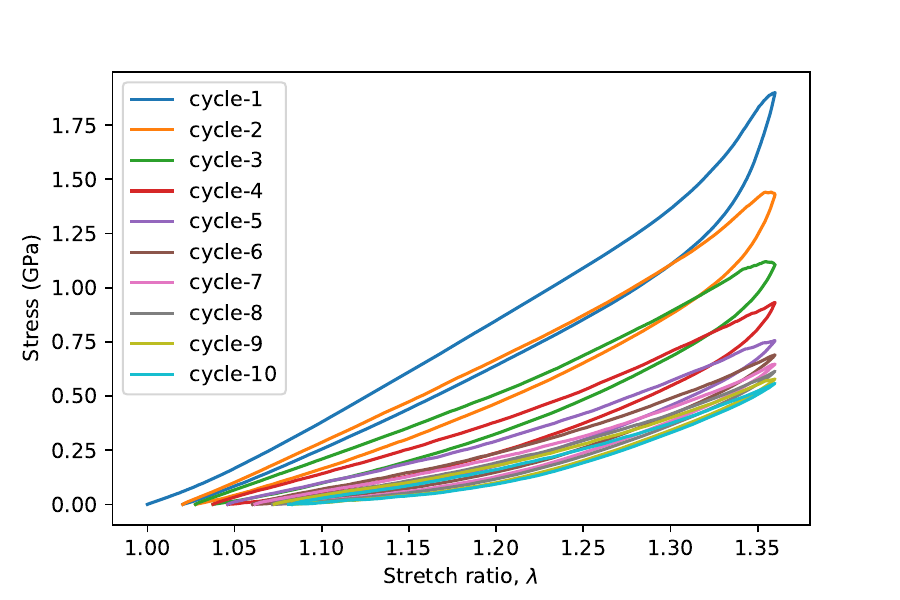}
	\caption {The stress-strain response of the collagen fibril model for strain controlled cyclic loading of a typical realization with $\beta=100\%$ and maximum stretch ratio $\lambda_{\max}=1.36$.}
	\label{fig4: ss_cyc_no_reform}
\end{figure}

We now examine the dissipative response of the fibril to cyclic loading for the case $\beta=100\%$, when all the end atoms are cross-linked to a neighboring atom. We perform cyclic loading simulation with maximum stretch ratio $\lambda_{\max}=1.36$, using the protocol described in Sec.~\ref{sim_prot}. We choose this particular  $\lambda_{\max}=1.36$ as it falls in the regime where cross-links are actively breaking (see Fig.~\ref{fig3: nb_mono_no_reform}). The fibril is subjected to $10$ loading-unloading cycles as in the experiment of Ref.~\cite{LIU2018cyclicload}. The stress-strain curves show the presence of dissipative hysteresis loops (see  Fig \ref{fig4: ss_cyc_no_reform}) in which the maximum stress as well as dissipation per cycle changes with increasing number of cycles. Furthermore, a residual strain is observed consistent with experimental data.  The associated hysteresis loops shift to the right with number of loading cycles showing accumulation of residual strain. These results reproduce features that were observed in cyclic loading experiment of fibrils~\cite{LIU2018cyclicload}.

\begin{figure}
	\centering
	\includegraphics[width = 1.0 \linewidth]{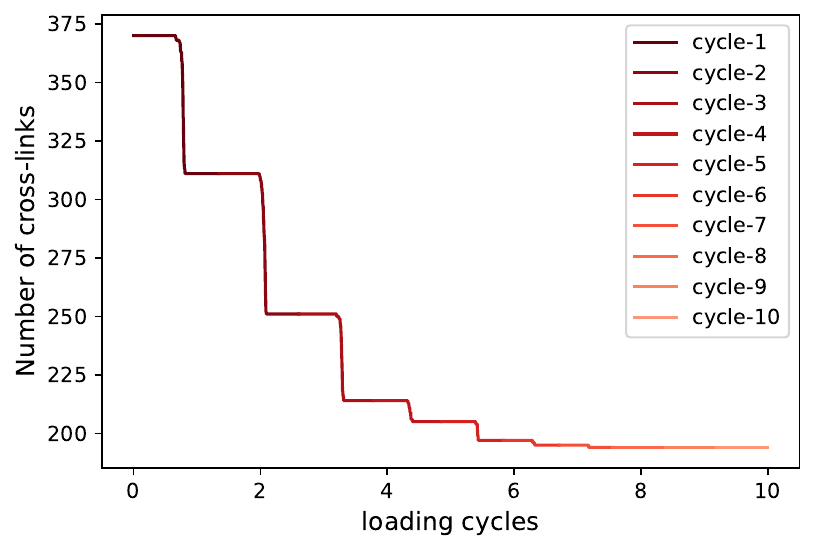}
	\caption {Time evolution of the number of cross-links under cyclic loading, corresponding to the stress-strain response shown earlier in Fig.~\ref{fig4: ss_cyc_no_reform}.}
	\label{fig5: nb_cyc_no_reform}
\end{figure}

To gain a insight into the microscopic mechanism behind the features of the hysteresis loops in Fig.~\ref{fig4: ss_cyc_no_reform}, we examine the associated time evolution of the number of cross-links, as shown in Fig.~\ref{fig5: nb_cyc_no_reform}. During the initial cycles, a higher fraction of the cross-links break. However, as the number of cycles increases, the rate of cross-link breakage per cycle decreases. This trend continues till the number of cross-links eventually approaches a steady state. The observed lowering of peak stress as well as accumulation of residual strain in Fig.~\ref{fig4: ss_cyc_no_reform} appears to be a direct outcome of the rupturing of the cross-links which make the fibril more compliant.

\begin{figure}
	\centering
	\includegraphics[width = 1.0 \linewidth]{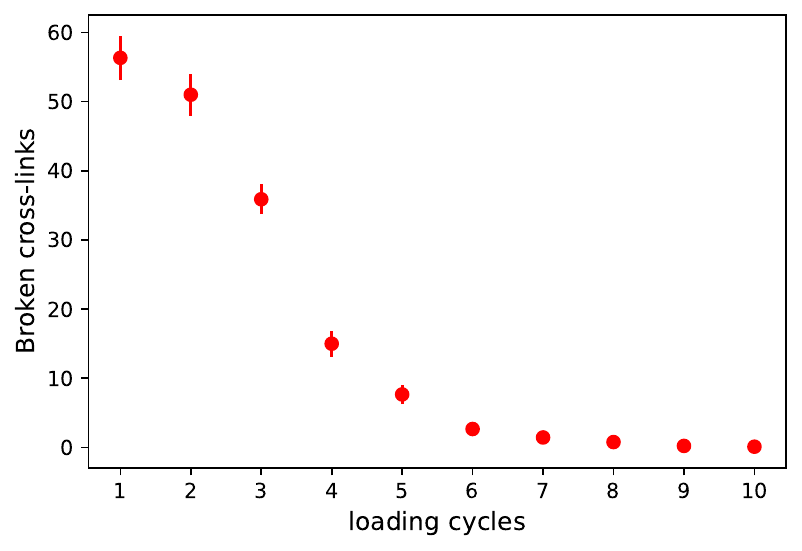}
	\caption {The average number of broken cross-links per cycle under cyclic loading, based on ten runs.}
	\label{fig6: nb1_cyc_no_reform}
\end{figure}

\begin{figure}
	\centering
	\includegraphics[width = 1.0 \linewidth]{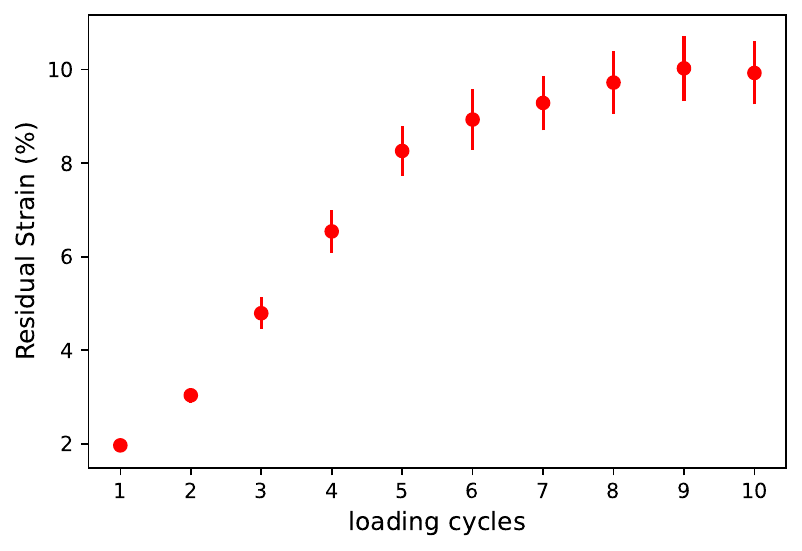}
	\caption {The evolution of residual strain with the number of loading cycles, averaged over ten runs.}
	\label{fig7: res_no_reform}
\end{figure}

\begin{figure}
	\centering
	\includegraphics[width = 1.0 \linewidth]{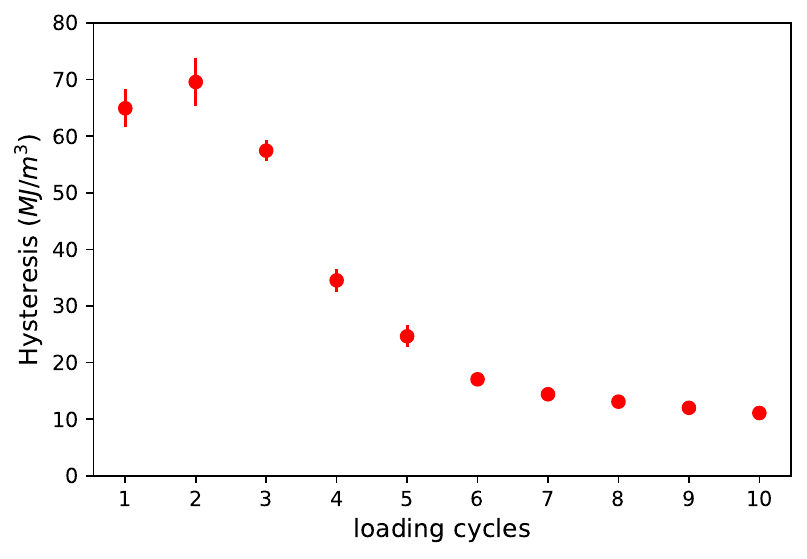}
	\caption {The evolution of energy dissipation with the number of loading cycles, averaged over ten runs.}
	\label{fig8: hys_no_reform}
\end{figure}

We next quantify the time evolution of the characteristic parameters of the stress-strain response to cyclic loading. To do so, we obtain the mean value of the parameters by averaging over 10 realizations. As seen in the typical realization earlier, the average number of number of broken cross-links per cycle also decreases with increasing number of cycles (see Fig.~\ref{fig6: nb1_cyc_no_reform}). Further, the associated residual strain, shown in Fig.~\ref{fig7: res_no_reform}, exhibits a corresponding increase to a steady state value of $\approx 10\%$ which falls within the range reported in experiment~\cite{LIU2018cyclicload}. The dissipation, evaluated as the area of the hysteresis loop, is seen in Fig.~\ref{fig7: res_no_reform} to have a marginal increase in the first two cycles before decreasing to a steady state value. The associated peak stress decreases monotonically with number of cycles to a steady state value, as seen in Fig. \ref{fig8: hys_no_reform}. 

\begin{figure}
	\centering
	\includegraphics[width = 1.0 \linewidth]{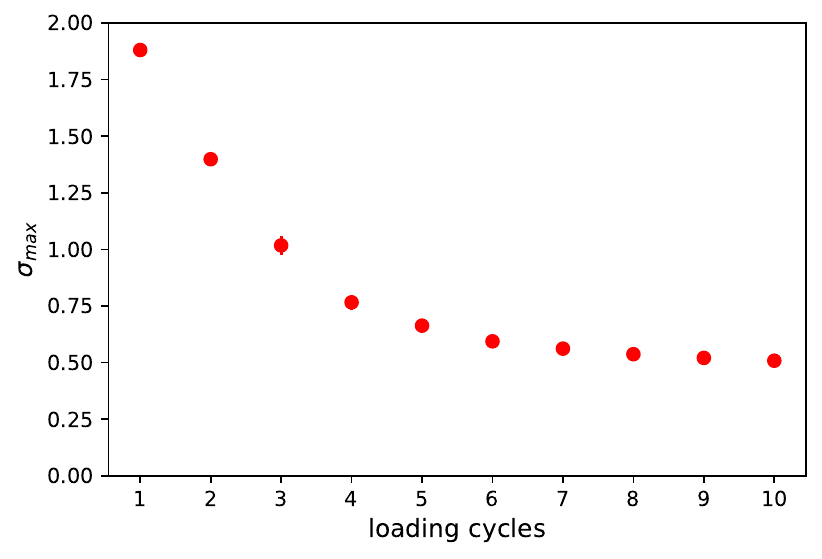}
	\caption {The variation of peak stress with the number of loading cycles, averaged over ten runs.}
	\label{fig9: sigma_no_reform}
\end{figure}

We now quantify the approach of the residual strain, energy dissipation, peak stress, and the number of broken cross-links per cycle to their corresponding steady-state values. As we found earlier for the kinetic model~\cite{suhail2022kinetic} and in the experimental data, we find that in the MD model too, the approach to the steady state is exponential with number of cycles $c$ (see Appendix~\ref{appendix:a}):
\begin{equation}
	q(c)-q(\infty) \propto e^{-c/c^*},
\end{equation}
where $q(c)$ represents the value of the relevant parameter after $c$ cycles. This allow us to extract the characteristic cycle number $c^*$. The value of $c^*$ is only weakly dependent on the choice of parameter for fixed $\lambda_{\max}$(see Appendix~\ref{appendix:a}).

\begin{figure}
	\centering
	\includegraphics[width = 1.0 \linewidth]{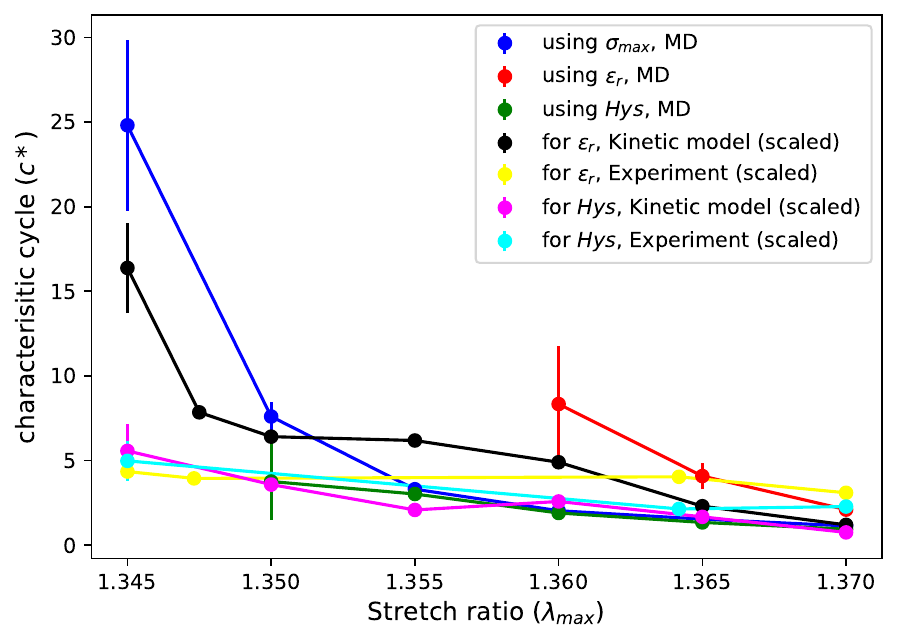}
	\caption {The variation of characteristic cycle, $c^*$ with the maximum stretch, $\lambda_{\max}$. For the MD model,$c^*$ extracted from $\sigma_{\max}$, residual strain and dissipation are shown. The stretch ratios for the kinetic model data and experimental data have been rescaled and are as reported in Ref.~\cite{ suhail2022kinetic}.}
	\label{fig11: c_vs_lambda}
\end{figure}

The characteristic cycle number $c^*$ depends on the choice of  $\lambda_{\max}$, as can be seen from Fig.~\ref{fig11: c_vs_lambda} where $c^*$ extracted from $\sigma_{\max}$, residual strain and dissipation are shown. If $\lambda_{\max}$ is close to the lower boundary of the region of stretches when cross-links break, then value of $c^*$ is relatively high. As $\lambda_{\max}$ is increased and lies within the range when cross-links break, then $c^*$ becomes independent of $\lambda_{\max}$ and is approximately $5$. The dependence of  $c^*$ on $\lambda_{\max}$ is also compared with that seen for kinetic model with sacrificial bonds~\cite{suhail2022kinetic} and experimental data in Fig.~\ref{fig11: c_vs_lambda}. We note that the ranges of stretches were different for kinetic model and experiment, and we have done a linear extrapolation to make the ranges coincide. We conclude that the results of MD are in good agreement with that of kinetic model as well as experiment. 

We further examine the dependence of $c^*$ on the extent of cross-linking $\beta$. We choose $\lambda_{\max}=1.345$ for which $c^*$ is relatively high for $\beta=100\%$. Cross-links are removed  to achieve the desired $\beta$. When $\beta$ is decreased, $c^*$ quickly decreases to a $\beta$-independent value which coincides with the value of $c^*$ for $\beta=100\%$ but higher $\lambda_{\max}$. This result can be understood by the rationale that while $\lambda_{\max}=1.345$ results in rupturing of only few cross-links for $\beta=100\%$, it lies well within the range of stretch ratios where significant number of cross-links break for smaller $\beta$.
\begin{figure}
	\includegraphics[width = 1.0 \linewidth]{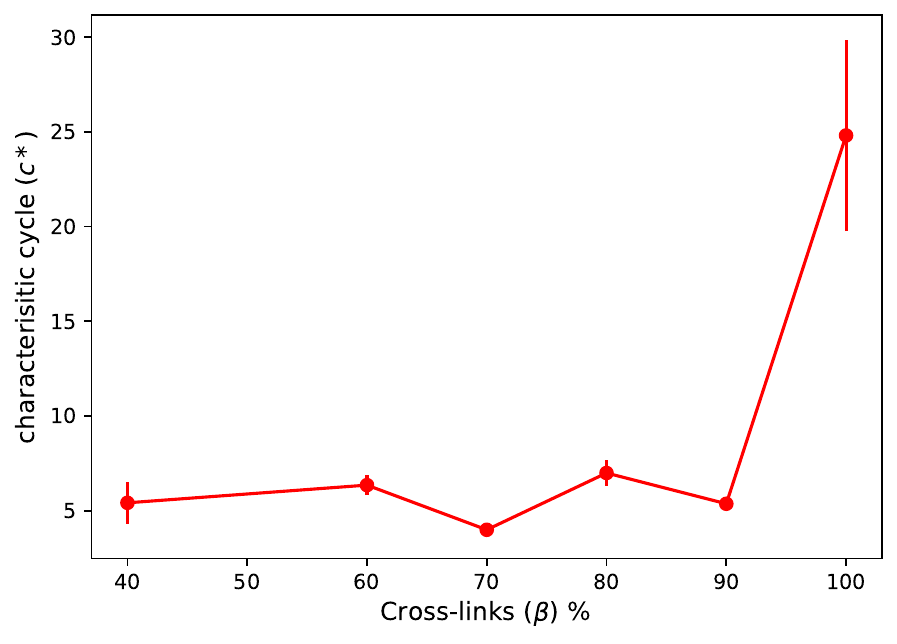}
	\caption {The variation of characteristic cycle, $c^*$, extracted from $\sigma_{\max}$, with the cross-link percentage, $\beta$, for  $\lambda_{\max}=1.345$. }
	\label{fig12: c_vs_beta}
\end{figure}

\begin{figure}
	\subfigure[]{\includegraphics[width=1.0\linewidth]{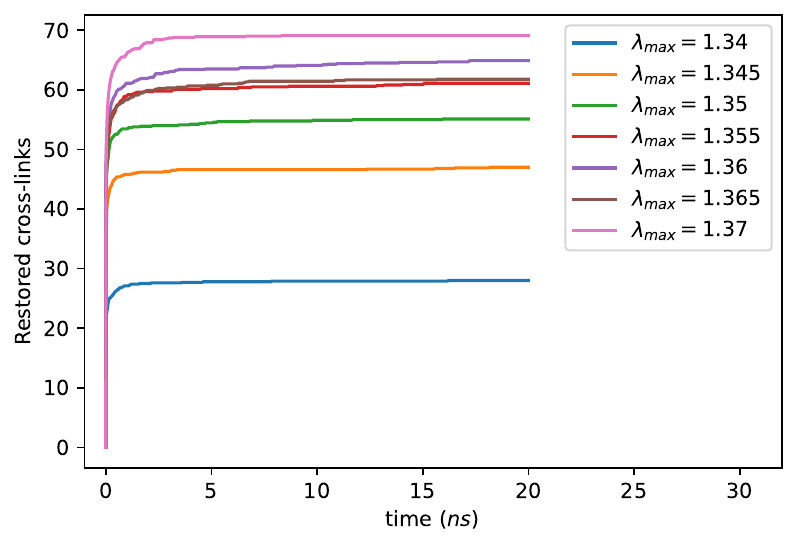}\label{fig:figure_b2}}
		\subfigure[]{\includegraphics[width=1.0\linewidth]{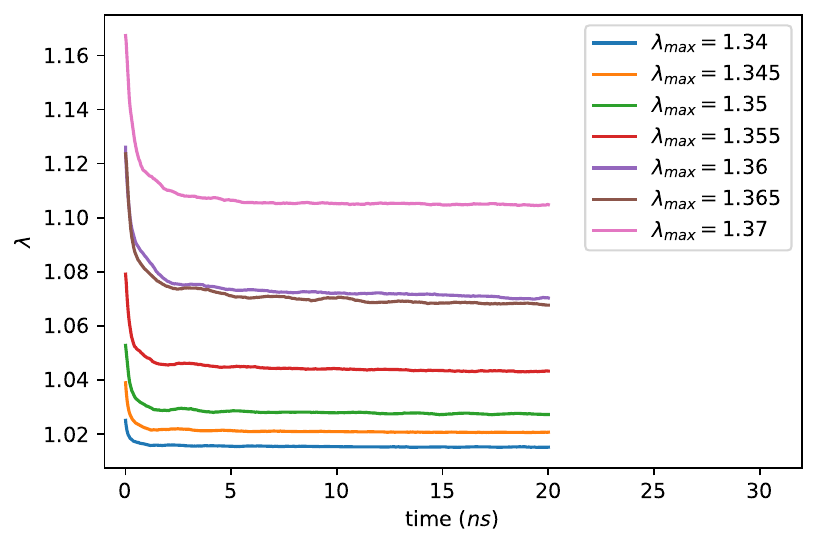}\label{fig:figure_a2}}
	\caption {Time evolution of (a) number of cross-links and (b) stretch ratio for  during relaxation at zero force after 10 cycles for different $\lambda_{\max}$. }
	\label{fig:relaxation}
\end{figure}

Having characterized the dissipation caused by cyclic loading, we now focus on the recovery when the fibril is allowed to relax at zero force after 10 cycles. Any pair of $E$ and $S$ atoms that approach less than a distance $r'=14$\AA form a cross-link instantaneously, provided neither atom is already part of any cross-link. The number of cross-links that reform increase with time till they saturate while the associated stretch ratio decrease with time to a steady state value, as can be seen in Fig.~\ref{fig:relaxation}(a) and (b) respectively. The saturation value of number of reformed cross-links increase with $\lambda_{\max}$ while that of residual strain decrease with $\lambda_{\max}$. We find that the recovery in strain in approximately $50\%$ for all $\lambda_{\max}$. This is comparable with the recovery seen in the experiment. We note that we could have introduced a time scale into the reformation process by associating a finite rate for the formation of cross-links. However, we find that saturation values are independent of the reformation rate. In the data shown in Fig.~\ref{fig:relaxation}, we set reformation rate to infinity as it is not practical to simulate relaxation for  60 minutes as in the experiment.

\begin{figure}
	\includegraphics[width = 1.0 \linewidth]{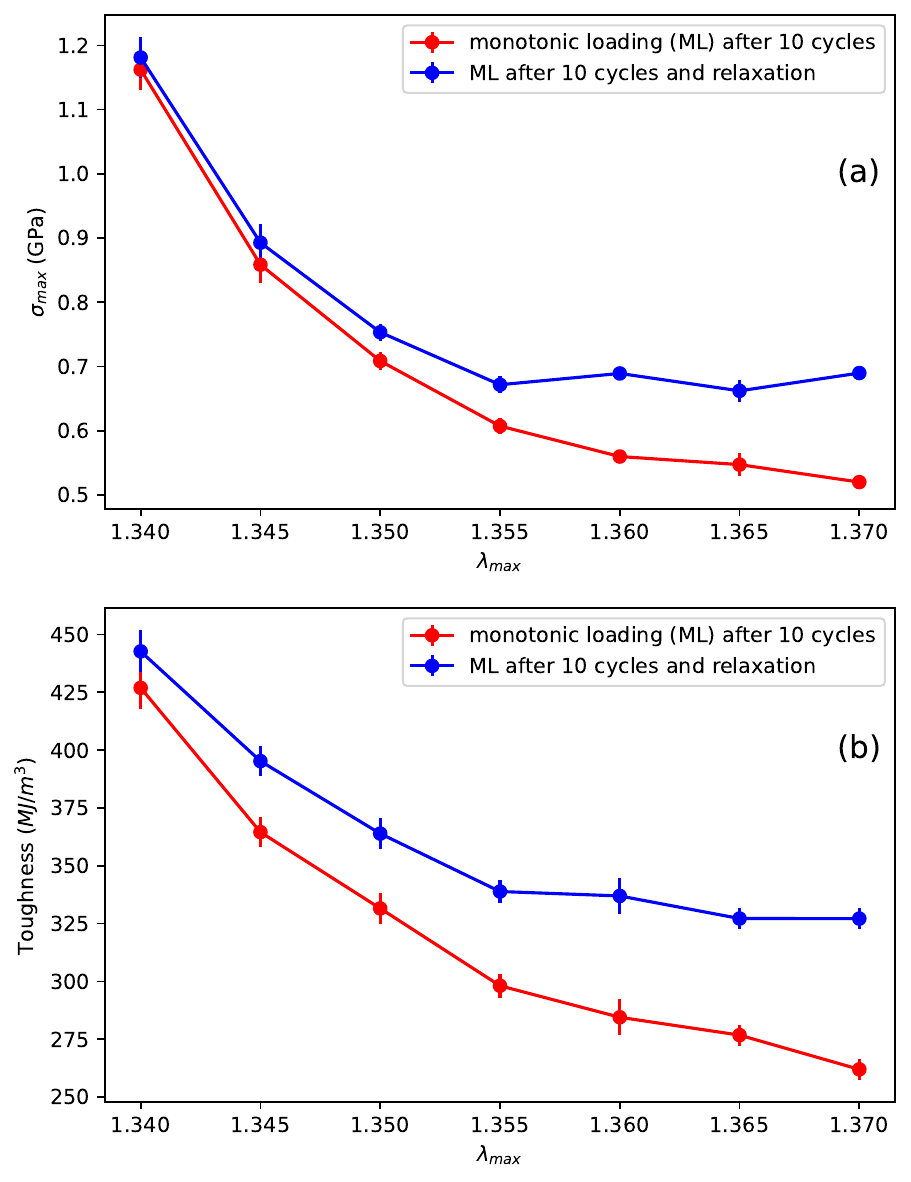}
	\caption {Comparison of (a) peak stress, $\sigma_{\max}$ and (b) toughness of a fibril subjected to monotonic loading with and without relaxation after being loaded for 10 cycles. Cross-links reform during the relaxation process.}
	\label{fig13: sigma_vs_lambda}
\end{figure}

To investigate the role of cross-link reformation on the macroscopic response of the fibril, we compare the following two cases. In case $1$, the fibril is directly subjected to monotonic loading after $10$ cycles. In case $2$, after $10$ cycles, we equilibrated the system for $20ns$ at zero force and then subjected it to monotonic loading. During the equilibration process, the cross-links were allowed to reform. We find that after relaxation (case $2$), the fibril shows increased strength and toughness (see Fig \ref{fig13: sigma_vs_lambda}). The difference in peak stress and toughness between the two cases become more significant as the maximum strain, $\lambda_{\text{max}}$, is increased. This can be attributed to the fact that at higher $\lambda_{\text{max}}$, a larger number of cross-links are  broken during the initial cyclic loading, resulting in more available free ends for cross-link reformation and recovery. It is important to note that while there is an increase in strength and toughness after the relaxation process, it does not exceed the original strength and toughness of the undamaged fibril. This observation indicates the presence of permanent plastic deformation resulting from cyclic loading, and it suggests that full recovery is not achievable within the framework of current model.

\begin{figure}
	\centering
	\includegraphics[width = 1.0 \linewidth]{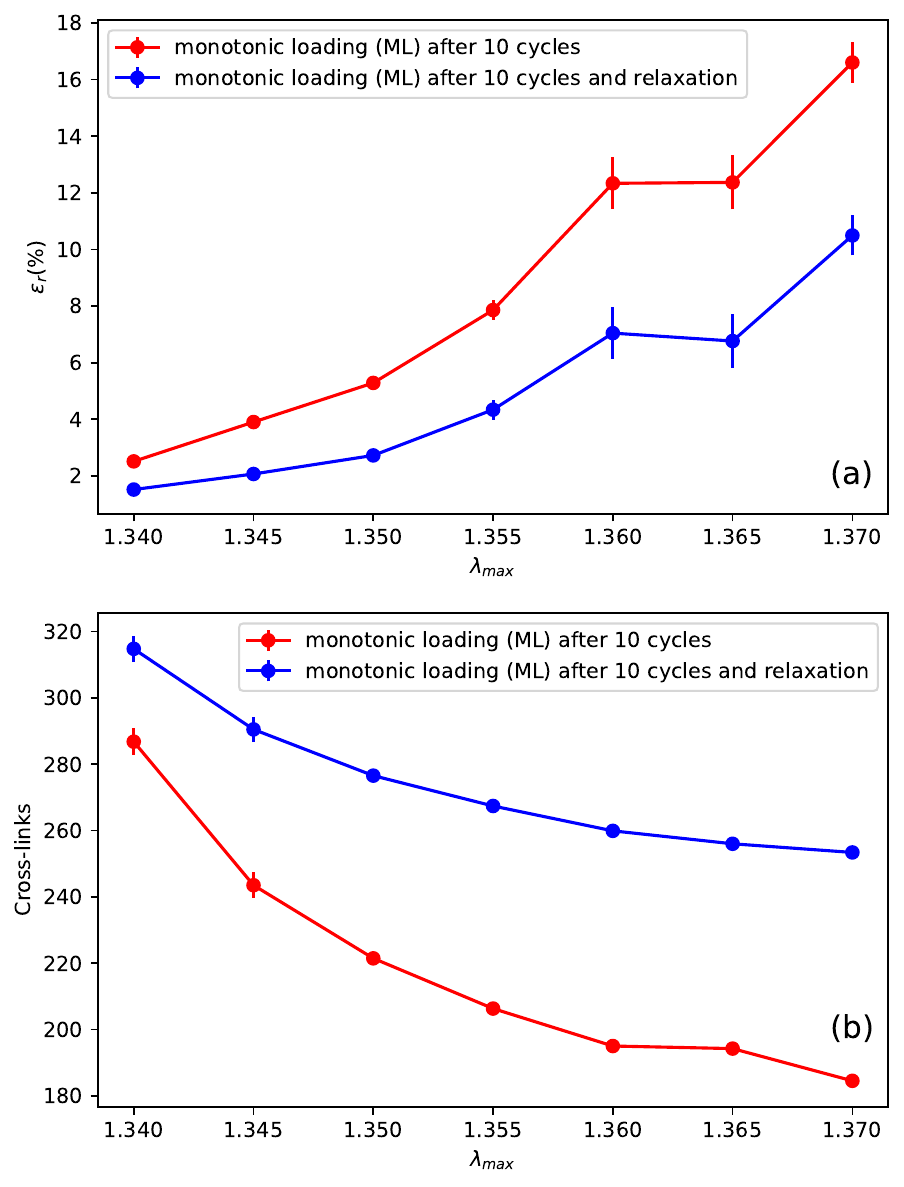}
	\caption{Comparison of (a) residual strain (b) total number of cross-links of a fibril subjected to monotonic loading with and without relaxation after being loaded for 10 cycles. Cross-links reform during the relaxation process.
}
	\label{fig14: nb_er_vs_lambda}
\end{figure}

Similar features -- improved characteristic parameters after relaxation -- can be seen for both residual strain and total number of cross-links, as shown in Fig.~\ref{fig14: nb_er_vs_lambda}. The change is more significant for larger $\lambda_{\max}$.

\section{Discussion and Conclusions}

Collagen, a widely present biomaterial, is of great importance, but there is currently less research focusing on fatigue experimental studies of individual collagen fibrils~\cite{SHEN2008cyclicload,SVENSSON2010,LIU2018cyclicload} when compared to the number of experiments on monotonic loading. In this paper, with emphasis on the experiment by Liu et. al~ \cite{LIU2018cyclicload}, we studied the dissipation and recovery of a collagen fibril when subjected to cyclic loads using molecular dynamics simulations of coarse grained models. Existing models for collagen fibrils, that have been obtained by coarse graining atomistic models, were earlier able to reproduce the macroscopic response to monotonic loading. Here, we incorporated reformation of cross-links or sacrificial bonds that aids in recovery. We show that the simulations reproduce key features of the cyclic loading experiment of Ref.~\cite{LIU2018cyclicload} such as moving hysteresis loops, residual strains, partial recovery on relaxation etc, and their dependence on different stretch ratios. The material parameters after relaxation were shown  improve with relaxation bringing out the role of extent of cross-linking in determining the macroscopic response.

The different parameters of the macroscopic response, such as peak stress, residual strain, dissipation, and number of cross-links approach the steady state values exponentially fast, characterized by a characteristic cycle number $c^*$. This behavior is consistent with what was observed in the analysis of the kinetic model~\cite{suhail2022kinetic} as well as seen in the experiment~\cite{LIU2018cyclicload}. We found that the $c^*$, becomes independent of the $\lambda_{\max}$, approximately equal to $5$, when $\lambda_{\max}$ lies within the regions where cross-links break, while it remains high at the lower boundary of this region. This observation is further supported by the dependence of $c^*$ on cross-link density $\beta$. Further, the value of $c^* \approx 5$ is same as that obtained for the kinetic model as well as in the experiment.

We investigated the post-cyclic loading recovery of the fibril model by allowing the fibril to relax and permitting cross-links to  reform during the relaxation process. We observe $\approx 50\%$ recovery in residual strain across different stretch ratios,  comparable with the results of the experiment~\cite{LIU2018cyclicload}. We do not find full recovery, thus there is plastic deformation. This is because cross-links form while the strain is reducing at zero force, thereby arresting further decrease in strain. Plastic deformation is consistent with the viscoelastic-plastic modeling approach of Ref.~\cite{fontenele2023understanding}, and the experimental results~\cite{LIU2018cyclicload}, but different from the kinetic model~\cite{suhail2022kinetic} where full recovery occurs if the fibril is relaxed for infinite time.  We note that this could be because the kinetic model is a minimal model that does not account for the complex geometrical structure of the collagen fibril.

To study the effect of cross-link reformation during relaxation, we compared the response to monotonic loading of two fibrils: one was subjected to monotonic loading immediately after cyclic loading, while the other was  relaxed and then subjected to monotonic loading. We observed an increase in strength and toughness in the fibril that underwent relaxation compared to the other one. However, this increase did not exceed the strength of the undamaged fibril under monotonic loading. This gain in strength, compared to no relaxation, is as in the viscoelastic-plastic model~\cite{fontenele2023understanding}. However, in the experiment of Liu et al.~\cite{LIU2018cyclicload}, the fibril that was directly subjected to monotonic loading immediately after cyclic loading, without relaxation, exhibited an increase in strength compared to the original fibril. This aspect is not reproduced neither in our molecular dynamics simulations, nor in the viscoelastic-plastic model~\cite{fontenele2023understanding} or kinetic model~\cite{suhail2022kinetic}. Understanding this phenomenon within models is a promising area for future study.

We also note that energy dissipation can show an increasing behavior with the loading cycles if the cyclic loading is done for small stretches ($\lambda_{\max}$), within the range where cross-links break. For example, energy dissipation increases until $3$ cycles for $\lambda_{\max}= 1.35$ before exhibiting an exponential decline (data not shown in the paper).

In the MD model, the range of stretch ratio ($\lambda_{\max}$) in which cross-links break is relatively narrow compared to both the kinetic model~\cite{suhail2022kinetic} and the experimental observations~\cite{LIU2018cyclicload}. The majority of cross-links break within the range $\lambda_{\max} \approx 1.34-1.37$.  In contrast, the kinetic model has a wider range for bond breaking, which is comparable to the experiment. This difference is due to the stochastic nature of the kinetic model where cross-links can break at different strain thresholds. In contrast, the cross-links in the MD model break at the same strain thresholds resulting in a narrower range of stretch ratios.

For the reformation, we could have introduced a new time scale in the form of rate of reformation. In the current study, we used an infinite rate, that is two atoms that are closer than the minimum distance form a bond instantaneously. Inclusion of a finite rate would slow down the reformation rate, but we have checked that the final number of cross-links is largely independent of this rate. Given that the relaxation times scale in the experiment is order of 60 minutes, our approach is justifiable provided we only analyze the steady state values and not the time-dependence.

The MD model that we studied has some limitations. Even though the MD model takes into account the three dimensional structure of the collagen fibril, it is still a simplification of the complex collagen fibril and it’s mechanics which also depends on several environmental factors like hydration etc. The parameters of these models are derived from atomistic simulation of small stretch of the collagen molecules with few bounded water molecules. The difference in parameters could arise because of intrinsic heterogeneity of the collagen molecule itself. Further, 
the stress-strain response and hence the value of the $c^*$ could depend on the distribution of the cross-link for $\beta <100\%$. One could do the same analysis done in this paper  including trivalent cross-links and combination of trivalent and divalent cross-links and/or Advanced Glycation Endproduct cross-links, which occur as a result of aging and diabetes~\cite{kamml2023influence}.

\begin{acknowledgments}

The simulations were carried out on the high performance computing machines Nandadevi at the Institute of Mathematical Sciences.

 \end{acknowledgments}

\appendix*
\section{\label{appendix:a} Approach of the characteristic parameters to steady state}

\begin{figure}
	\includegraphics[width = 1.0 \linewidth]{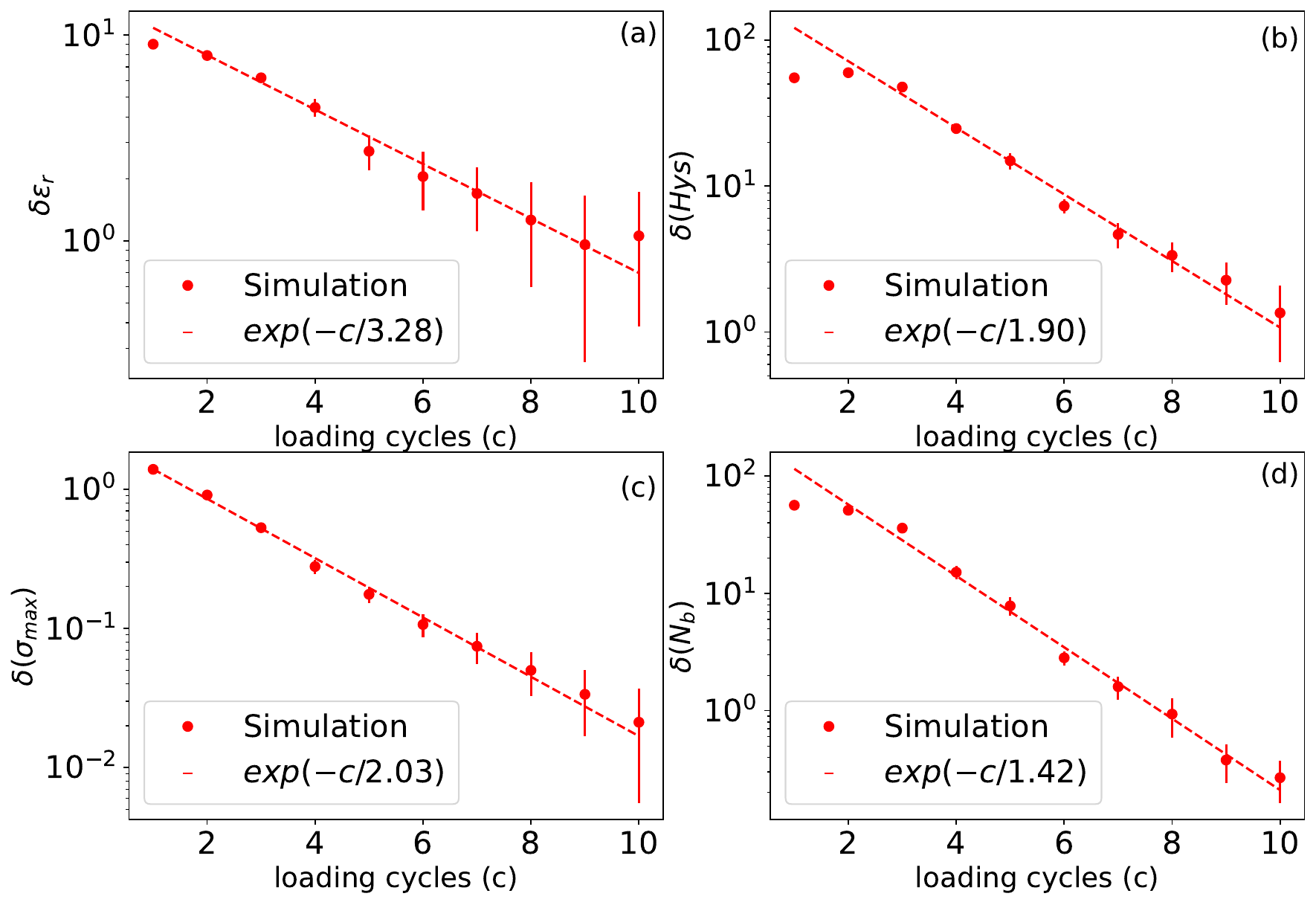}
	\caption {The  (a) residual strain, (b) hysteresis, (c) peak stress and (d) number of broken bonds per cycle approach their respective steady state values exponentially fast. The data are for $\lambda_{\max}=1.36$.}
	\label{fig10: fit-res-hys-sigma-nb}
\end{figure}

We find that the differences of residual strain, energy dissipation, peak stress, and broken bonds per cycle from their steady state values decrease exponentially to zero with the number of loading cycles, as can be seen in Fig.~\ref{fig10: fit-res-hys-sigma-nb}. 

%

\end{document}